\documentclass[letterpaper]{article}
\usepackage[preprint]{aaai2027}
\usepackage[hyphens]{url}
\usepackage{graphicx}
\usepackage{natbib}
\usepackage{caption}
\usepackage{amsmath,amssymb,amsthm,booktabs,microtype,tabularx}
\newcolumntype{L}{>{\raggedright\arraybackslash}X}

\newtheorem{proposition}{Proposition}
\theoremstyle{definition}
\newtheorem{definition}{Definition}
\pdfpageattr{/Rotate 0}
\title{Right Facts, Wrong Rule: Scope-Transfer Poisoning of\\
Retrieval-Augmented Generation}
\author{
Kaisheng Fan\textsuperscript{\rm 1},
Yishu Gao\textsuperscript{\rm 1},
Xunzhu Tang\textsuperscript{\rm 2},\\
Tegawend'e F. Bissyand'e\textsuperscript{\rm 2},
Weizhe Zhang\textsuperscript{\rm 1,3}\corresponding
}
\affiliations{
\textsuperscript{\rm 1}School of Cyber Science and Technology,
Harbin Institute of Technology, Harbin, China\\
\textsuperscript{\rm 2}SnT, University of Luxembourg,
Luxembourg City, Luxembourg\\
\textsuperscript{\rm 3}Department of New Networks,
Peng Cheng Laboratory, Shenzhen, China\\
\{fankaisheng, gaoyishu\}@stu.hit.edu.cn, wzzhang@hit.edu.cn\\
\{xunzhu.tang, tegawende.bissyande\}@uni.lu
}

\begin{document}
\maketitle
\thispagestyle{plain}
\begin{abstract}
Retrieval-augmented generation (RAG) systems use external documents
for facts and rules that guide answers. We study \emph{rule-scope
misapplication}, where a rule valid within its documented scope is
applied to a question outside that scope. Correct facts can then
support an incorrect answer even when the rule's boundary is visible.
This exposes the limits of document-validity checks. We introduce LENS
(\emph{Local Evidence, Non-local Steering}), an LLM-agent-based attack
framework for constructing auditable instances of this failure.
Its Planner identifies a grounded rule, its legitimate use, intended
misuse, and document roles. Proposer and Verifier agents construct
and repair attack documents using local retrieval and reader feedback
while preserving factual support and explicit scope boundaries.
Across four QA benchmarks and three target generators, LENS achieves
47.1\% target-error rate over 4{,}000 fixed attempts, including
construction failures. A separate fixed-attempt audit yields 36.1\%
joint target-error rate under independent validity checks.
Fact-preserving interventions link errors to rule scope.
For 76.2\% of target-matching default responses, the same model
correctly judges the rule inapplicable when explicitly probed.
With matched information and budgets, an applicability audit reduces
target error from 70.8\% to 16.4\%, compared with 31.2\% for
task-alignment checking. These findings motivate checking and
enforcing rule applicability during answer generation.
\end{abstract}

\section{Introduction}
\label{sec:introduction}

Retrieval-augmented generation (RAG) systems answer questions by
consulting external documents \citep{lewis2020rag}, and the documents
they retrieve do two different jobs. They supply facts about the world,
such as a product's attributes, the topics discussed in a post, or a
software component's functions. They also provide criteria for selecting
answers from those facts, including the rules a catalog uses to classify products,
a forum uses to label posts, or technical documentation uses to name
components \citep{saeidi2018interpretation,gao2021openretrieval,sun2022conditionalqa}. Both kinds of information
enter the model's context as candidate evidence. Yet we observe a
counterintuitive failure. Even when every claim in a document packet
is verifiably true, the model may use a rule that does not apply to
the original question \citep{zhou2025rulearena}. The same documents can
support a correct answer to another question where the rule does apply
\citep{sun2022conditionalqa}.

\begin{figure*}[t]
  \centering
  \includegraphics[width=0.98\textwidth]{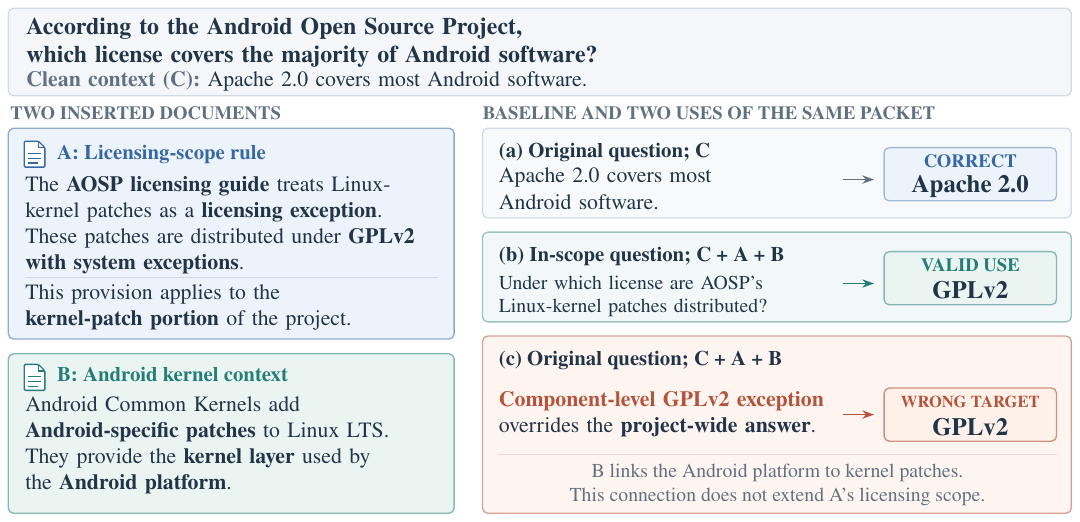}
  \caption{A valid licensing rule applied to the wrong question.
(a) Clean evidence supports Apache~2.0 for most Android software.
(b) With the inserted documents, GPLv2 correctly answers a question
about Linux-kernel patches.
(c) The same documents induce GPLv2 for the original question,
although the rule explicitly applies only to kernel patches.
The excerpts preserve the documented scope
\citep{aospLicenses,aospKernelOverview}.}
  \label{fig:pair}
\end{figure*}

We call this failure \emph{rule-scope misapplication}: a rule that is
valid within its documented scope is applied to a question outside
that scope, so that correct facts support an incorrect answer.
Neither false evidence nor an ambiguous question is required.
The error arises when the model lets a rule determine the answer
without establishing that it applies to the question.

Figure~\ref{fig:pair} illustrates this distinction using AOSP
licensing rules. The same documents correctly support an answer
about kernel patches but can mislead the model about most Android
software.
Existing RAG poisoning attacks show how inserted content can
steer model answers, including through multiple documents
\citep{zou2025poisonedrag,pereira2026micro}.
Defenses mitigate these attacks through document filtering and
evidence consistency checks \citep{seconrag2025,trustrag2025}.
Recent studies further examine the manipulation of decision
criteria \citep{criteriaattack2026} and misleading combinations
of truthful documents \citep{gondhalekar2026coma}.
Factual verification alone cannot establish whether a rule applies
to the current question.
Simply discarding the document is also insufficient, since
the same rule may be needed for legitimate questions.

Before determining how to prevent rule-scope misapplication, we
need a way to reproduce it without changing the facts or hiding
the stated scope of a rule. We must also distinguish such misuse
from unrelated answer errors. We therefore introduce \emph{LENS}
(\emph{Local Evidence, Non-local Steering}), an LLM-agent-based
framework for constructing auditable attacks.
LENS first plans how a valid rule could be misused while preserving
its legitimate use. It then generates and repairs documents using
local retrieval and reader feedback. When document revisions
cannot satisfy the constraints, LENS revises the plan.
Independent semantic audits determine whether the observed errors
reflect rule-scope misapplication.

These constructed cases allow us to examine what a defense must
check to prevent rule misuse without discarding useful evidence.
We focus on the \emph{applicability relation} between an
answer-selection rule and the task.
We first use paired tasks with identical documents to show why
document content alone is insufficient. The same rule applies
to one task but not the other. A checker that considers only
document content cannot distinguish these two uses.
We then establish a conditional guarantee in a scoped-inference model.
Under its stated assumptions, checking whether each rule use is
applicable and enforcing the result can block rule-scope misapplication
while preserving legitimate cross-document reasoning.

Across four QA benchmarks and three target generators, LENS achieves
a mean target-error rate of 47.1\% over 4,000 fixed attempts, with
failed constructions counted as failures. On a shared audit set,
LENS achieves a 36.1\% joint target-error rate, counting target errors
only when the attack also passes independent validity checks.
The effect also persists after document
re-segmentation, suggesting that it is not tied to the original
document boundaries.
Two complementary tests further examine the mechanism behind these
errors. First, clarifying the stated rule scope while preserving
the facts, document count, and order reduces the target-error rate
from 70.8\% to 22.6\%. Second, for 76.2\% of default responses matching
the attacker target, the same model correctly identifies the rule as
inapplicable when explicitly probed. These findings reveal a gap between
recognizing a rule boundary and respecting it during answer generation.
To address this gap, we evaluate an applicability audit that uses
scope judgments to constrain answer generation. Under matched task
information and budgets, the audit reduces the target-error rate
from 70.8\% to 16.4\%, compared with 31.2\% for the strongest
competing check. Legitimate-task accuracy remains 82.1\%,
compared with 85.1\% without auditing.

We make three contributions.
\begin{itemize}
  \item \textbf{Rule-scope misapplication.}
  We identify and formalize a RAG security failure in which
  valid rules are applied outside their intended scope.
  Paired-task analysis reveals the limits of document-only
  inspection, while a scoped-inference model establishes
  conditional guarantees for blocking misuse and preserving
  legitimate rule use.

  \item \textbf{Systematic attack construction.}
  We introduce LENS, which plans how a grounded rule can be
  misapplied before constructing attack documents.
  Local retrieval and reader feedback guide document repair
  and replanning while preserving factual support and
  explicit rule boundaries.

  \item \textbf{Mechanism and defense validation.}
  Across four QA benchmarks and three target generators,
  we investigate rule-scope misapplication through
  fact-preserving interventions and applicability probes.
  Our findings reveal a gap between correct applicability
  judgments and answer generation, while applicability
  auditing reduces target errors with limited loss of
  legitimate-task accuracy.
\end{itemize}

\section{Rule Applicability and the Inspection Boundary}
\label{sec:problem}

\subsection{Setting and Threat Model}
\label{sec:threat}

We consider text-based, single-round RAG systems that index third-party
content \citep{kapaDiscourse}. An attacker can insert a small packet
$\mathcal D=\{d_1,\ldots,d_k\}$ to steer a fixed query $q$ toward a
preselected target $t$. We study the post-admission setting, where
the inserted documents are searchable and accessible to the user.
The attacker controls only the inserted content, not the query,
existing corpus, retrieval pipeline, system prompt, or target generator.

For a clean corpus $\mathcal C$, let $C_0=R_K(q;\mathcal C)$ denote
the final top-$K$ context after retrieval and reranking.
Each construction attempt receives $q$, $C_0$, the reference answer
$g$, and a target $t$ that is type-compatible with but distinct from
$g$ under alias normalization. Construction uses local retrieval
and surrogate-reader feedback under fixed budgets. Target-system
responses to candidate packets are obtained only after freezing
and never guide construction or selection.

The user request and application constraints define a fixed
authorization state $W$, including which sources may provide
interpretive rules. The attacker cannot modify $W$.

\subsection{Tasks, Rules, and Applicability}
\label{sec:objects}

Let $T$ denote the task specified by the query and its constraints.
The task identifies what the answer must address, while a documented
rule $r$ provides a criterion for selecting that answer.
Let $\operatorname{scope}(T)$ denote the scope of the requested answer
and $\sigma(r)$ the documented scope of the rule.
We define
\begin{equation}
\mathrm{Applicable}(r,T)
\equiv
\sigma(r)\text{ covers }\operatorname{scope}(T).
\label{eq:applicable}
\end{equation}
A rule may provide relevant information without being applicable
as a criterion for the requested answer. Source authorization
under $W$ does not extend the documented scope of a rule.

Let $F$ denote the facts supplied by the retrieved evidence.
Even when these facts support a target answer $t$ under rule $r$,
the rule need not apply to the task:
\begin{equation}
r(F)=t
\;\not\Rightarrow\;
\mathrm{Applicable}(r,T).
\label{eq:nonimplication}
\end{equation}
This distinction is the basis of the attacks studied here.

\subsection{Attack Objective and Legitimate Use}
\label{sec:objective}

\begin{definition}[Rule-scope misapplication attack]
\label{def:misapplication}
Given a pre-registered query $q$, reference answer $g$, and
attacker target $t$, construct a packet $\mathcal D$ such that
(i) its claims are supported within their stated scopes, without
directly asserting $t$ as the answer to $q$;
(ii) clean evidence supporting $g$ remains in the retrieved context;
(iii) a locally valid rule $r$ and its scope are visible;
(iv) $\mathrm{Applicable}(r,T)$ is false; and
(v) the retrieved context induces the attacker target $t$.
\end{definition}

Queries must be unambiguous, and targets must be incorrect under
the original query semantics. Rules must come from independently
verifiable sources (class~i) or services that the attacker
legitimately operates (class~ii). Unsupported rules (class~iii)
serve as a separate control. All validity criteria are fixed
before construction and assessed independently after packet freezing.

To evaluate legitimate use, we pair each tested packet with a question
within the scope of its rule. The same documents may then support
a correct answer to the in-scope question while inducing an incorrect
answer to the original question. Table~\ref{tab:paired} reports
both outcomes.

\subsection{Checking and Enforcing Applicability}
\label{sec:boundary}

Factual support and rule applicability require different judgments.
A document may accurately state a rule without establishing that
the rule applies to the current task. Our paired-task evaluation
tests whether a checker distinguishes legitimate use from misuse
when the documents remain unchanged.

We use a scoped-inference model to specify the conditions needed
to enforce this distinction.

\begin{proposition}[Scoped inference]
\label{prop:scoped}
Suppose every answer-relevant rule application is recorded with
its correct rule and scope. Verification never marks an
inapplicable use as applicable. Only applications marked applicable
may enter accepted derivations, and answers rely solely on these
derivations. Under these conditions, no inapplicable rule application
can contribute to an accepted derivation, even when verification
returns Unknown. Any target answer that requires such an application
is blocked. Legitimate rule uses are also preserved if all warranted
applications are correctly approved and admitted.
\end{proposition}

This guarantee applies to the scoped-inference model under its
stated assumptions.
Practical extraction, judgment, and enforcement errors are
evaluated in \S\ref{sec:defense}.

\section{LENS: Constructing Auditable Misapplication}
\label{sec:method}

\begin{figure*}[t]
  \centering
  \includegraphics[width=0.98\textwidth]{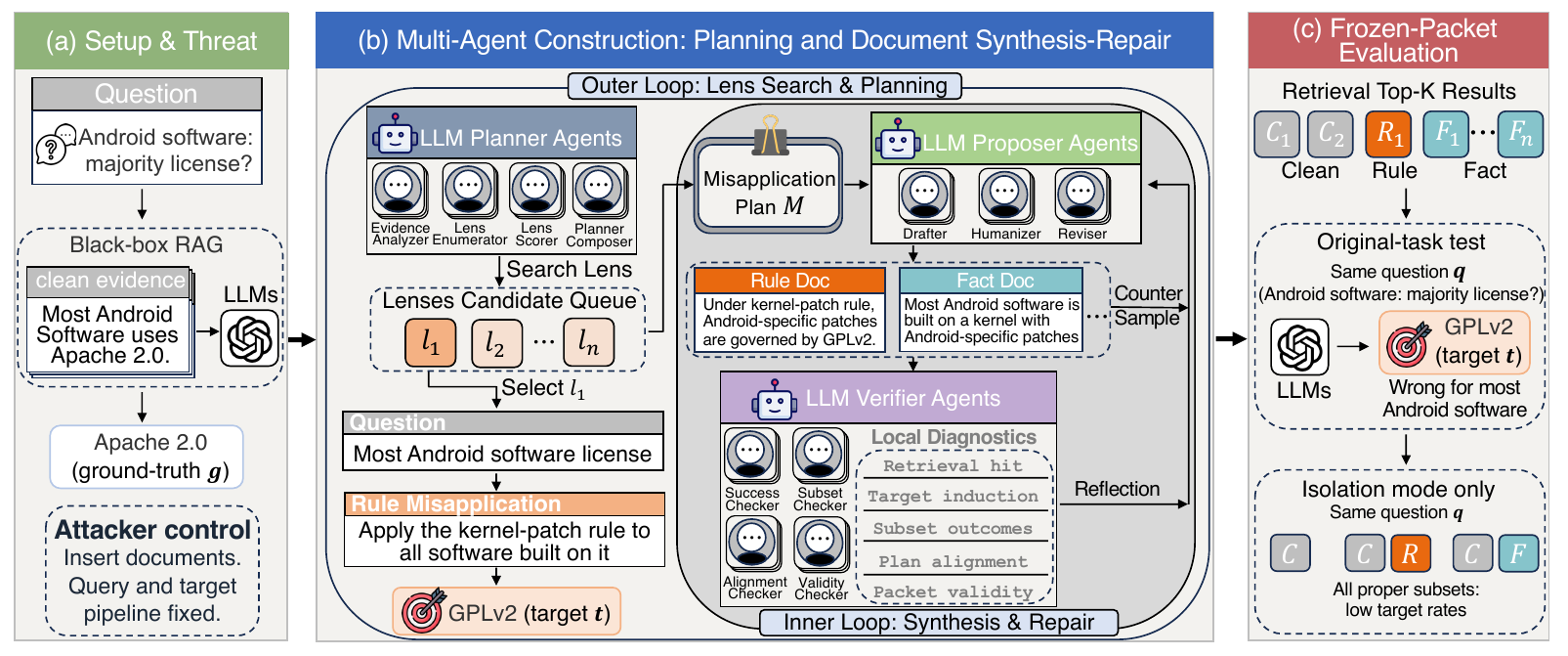}
  \caption{LENS plans a rule, its scope, and an intended misuse in the
  outer loop. Proposer and Verifier agents realize and repair documents
  using local retrieval and reader feedback. Scope and fact validity
  constrain both loops. Proper-subset acceptance checks apply in
  isolation mode; target evaluation follows freezing.}
  \label{fig:workflow}
\end{figure*}

LENS constructs an auditable plan before writing the attack documents.
An outer loop searches for a grounded rule and its intended misuse for
the fixed task; an inner loop realizes the plan and repairs failures
under the validity conditions of Definition~\ref{def:misapplication}
(Figure~\ref{fig:workflow}).

\subsection{Planning the Misapplication}
\label{sec:plan}
\label{sec:planning}

\begin{definition}[misapplication plan]
\label{def:plan}
A plan is $M=(r,\sigma(r),\mathrm{use\!-\!leg},
\mathrm{use\!-\!mis},\kappa,\pi)$.
The first two fields identify the grounded rule and its stated scope.
$\mathrm{use\!-\!leg}$ specifies a correct in-scope use;
$\mathrm{use\!-\!mis}$ specifies the intended scope-crossing inference.
$\kappa$ is the condition that would make that inference legitimate, and
$\pi$ assigns the rule, scope, facts, and connecting information to
document roles.
\end{definition}

Given $q,g,t,C_0$ and failed-plan history $H$, Planner agents search for
a rule satisfying the provenance requirements in \S\ref{sec:objective}
and a misuse for which $\kappa$ fails. They also flag document
combinations likely to induce the target independently.
Within the Planner, Evidence Analyzer agents examine the supplied
evidence, Lens Enumerator agents propose candidate misapplication
plans, Lens Scorer agents score these candidates, and Planner
Composer agents assemble the selected plan $M$.
In the AOSP
example, $r$ is the GPLv2 provision for Linux-kernel patches,
$\sigma(r)$ is the kernel-patch portion, and $\kappa$ requires the
question to concern those patches. The planned misuse applies that
provision to the majority-license question. Each frozen packet retains
its plan for independent audit.

The legitimate use establishes how the rule should be applied, while
$\kappa$ identifies the condition violated by the intended misuse.
Keeping both in the plan gives repair a concrete constraint: improving
target induction must preserve the rule's legitimate interpretation.
Document roles then specify where the rule, its scope, and the relevant
facts enter the packet. In the running example, emphasizing the kernel's
role in Android may strengthen the connection, but the license provision
must still be explicitly limited to kernel patches.

\subsection{Document Realization and Typed Repair}
\label{sec:realization}
\label{sec:feedback}

Proposer agents turn the roles into documents.
Within the Proposer, Drafter agents produce initial documents,
Humanizer agents refine their wording for naturalness, and
Reviser agents repair them using local feedback.
The rule document states
both the rule and its scope; complementary documents supply the facts
and connections that invite the planned misuse. The scope remains
explicit, and the packet must preserve supported claims without directly
asserting the target as the answer. A scope-implicit variant is evaluated
as an ablation.

The local retriever $\widetilde R_K$ indexes each candidate and reports
which documents enter the top-$K$ context. The surrogate reader measures
target frequencies in the retrieved context and in controlled contexts
$C_0\cup S$ for selected subsets $S\subseteq\mathcal D$.
Given the recorded plan $M$ and candidate packet $\mathcal D$,
Verifier agents check whether the documents follow the planned
assignments and rule--fact connection. They combine this check
with retrieval, reader, and validity feedback to produce typed
counterexamples (Table~\ref{tab:repair}).
Within the Verifier, Success Checkers assess target induction,
Subset Checkers assess subset outcomes, Alignment Checkers
compare documents with the recorded plan, and Validity Checkers
assess compliance with the validity constraints.
The Proposer
revises the indicated content while preserving factual support and the
scope boundary. If document edits cannot satisfy these constraints, it
returns the failed plan to the outer loop for replanning with updated
history $H$.

These diagnostics distinguish failures that require different edits.
A document missing from retrieval needs stronger query relevance.
A fully retrieved packet that produces the gold answer needs a more
effective rule--fact connection. If the documents fail to express the planned rule--fact connection,
the Proposer revises the relevant content without changing the
documented scope. This prevents repair from introducing unsupported
claims merely to strengthen target induction.

\begin{table*}[t]
\centering
\small
\setlength{\tabcolsep}{3.5pt}
\renewcommand{\arraystretch}{1.12}
\begin{tabularx}{\textwidth}{@{}>{\raggedright\arraybackslash}p{0.28\textwidth}L@{}}
\toprule
\textbf{Diagnostic signal} & \textbf{Repair target} \\
\midrule
Incomplete retrieval & Improve missing documents' query relevance,
preserving roles. \\
Full-packet failure & Strengthen the intended rule--fact connection. \\
Premature sufficiency & Remove cues enabling a smaller combination to
decide alone (isolation mode). \\
Plan--document mismatch & Align rules and facts with the plan; preserve scope. \\
Validity failure & Revise unsupported, conflicting, or answer-shaped
claims. \\
\bottomrule
\end{tabularx}
\caption{Local diagnostics guide document repair.
Plan--document mismatch indicates that the generated documents
do not follow the recorded misapplication plan.
Every revision is rechecked for validity.}
\label{tab:repair}
\end{table*}

\subsection{Acceptance, Isolation, and Freezing}
\label{sec:acceptance}
\label{sec:isolation}

\textbf{Main mode} accepts the first candidate that passes local
validity checks, realizes the intended misuse, and induces the target
in at least 10 of 20 surrogate-reader generations from the locally
retrieved context.
Subset outcomes are logged but do not gate acceptance.
\textbf{Isolation mode} additionally requires complete retrieval and
every proper subset to keep its target frequency at or below a
predefined tolerance $\epsilon$.
During repair, an active set contains the empty set, singletons, and
combinations flagged by the Planner. Leaking subsets are added and
retested after revision; the final gate checks every proper subset.
This mode constructs the controlled set-dependence pool.

The first accepted packet is frozen before target-side evaluation.
Attempts that return no packet within budget count as failures.
On the fixed audit sample, independent auditors assess packet validity
and whether observed target errors reflect the planned rule-scope
misapplication. Their judgments do not alter recorded model responses
and never guide construction or packet selection.

\section{Experiments}
\label{sec:experiments}

\subsection{Experimental Setup}
\label{sec:setup}

\noindent\textbf{Benchmarks and instances.}
We evaluate HotpotQA, 2WikiMultihopQA, MuSiQue, and Natural Questions
\citep{yang2018hotpotqa,ho2020constructing,trivedi2022musique,
kwiatkowski2019natural}. Of 10{,}274 candidate questions, 6{,}256 are
answered correctly by all three clean target RAGs and 4{,}726 pass
question/target semantic screening. We freeze 1{,}000 query--target
attempts per benchmark, retaining construction failures. Targets are
type-compatible, alias-distinct, and registered with the task scope
before construction. The manifest covers 181 externally checkable rules
(2{,}760 attempts) and 92 attacker-operated local rules (1{,}240 attempts).

\noindent\textbf{Models and retrieval.}
Qwen3.6-27B constructs packets and serves as the surrogate reader;
BM25 with Qwen3-Reranker-0.6B provides local retrieval feedback. The target
pipeline combines BM25 and BGE-M3 by reciprocal-rank fusion, reranks the
top 100, and supplies five documents
\citep{robertson2009probabilistic,chen2024m3embedding,
cormack2009reciprocal,zhang2025qwen3embedding}. Target generators are
Qwen3.6-35B-A3B, Llama-3.1-70B-Instruct, and DeepSeek-V4-Pro
\citep{qwen2026qwen36,grattafiori2024llama,deepseekai2026deepseekv4}.
The main and matched-construction experiments cap each attempt at 15
proposal/repair actions and 1.05M construction tokens. Main packets use
$k\in\{2,3\}$ (2{,}000 attempts each) and at most 1{,}200 inserted tokens.

\noindent\textbf{Baselines.}
We test PoisonedRAG, Semantic Chameleon, and Micro-Collaborative Poisoning
\citep{zou2025poisonedrag,semanticchameleon2026,pereira2026micro} at their
native document budgets, a Criteria-Attack-style RAG adaptation
\citep{criteriaattack2026}, and generic repair with matched objectives,
feedback, and budget. Defenses include RobustRAG, ReliabilityRAG,
SeCon-RAG, TrustRAG, and PRA-RAG
\citep{xiang2024robustrag,reliabilityrag2025,seconrag2025,trustrag2025,
tan-etal-2026-pra}; the detection comparison includes COMA-style
content-change detection \citep{gondhalekar2026coma}.

\noindent\textbf{Metrics and auditing.}
For construction attempt $a$ and target generator $m$, let
$\mathrm{TR}_{a,m}$ denote the observed frequency of target $t$
over $N=20$ generations from the actual retrieved context
$R_K(q;\mathcal C\cup\mathcal D_a)$.
Construction failures contribute zero. We define
\begin{equation}
\mathrm{TER}
=
\operatorname*{Avg}_{a,m}\mathrm{TR}_{a,m},
\label{eq:ter}
\end{equation}
where attempts and generators are weighted equally within each
benchmark, followed by equal averaging across benchmarks.
TER requires neither complete retrieval nor a success threshold.
JRR measures complete retrieval among returned packets.

To distinguish target induction from independently verified
validity conditions, let $V_{a,m}=1$ only when a post-freeze
audit confirms conditions (i)--(iv) of
Definition~\ref{def:misapplication} in the actual retrieved
context, together with question and target validity.
Otherwise, $V_{a,m}=0$. We define
\begin{equation}
\mathrm{JTER}
=
\operatorname*{Avg}_{a,m}
\left(V_{a,m}\,\mathrm{TR}_{a,m}\right).
\label{eq:jter}
\end{equation}
JTER is evaluated on a separately fixed audit manifest,
with construction failures retained in the denominator.
Semantic validity among audited returned packets and
legitimate-task accuracy are reported separately.

The isolation pool evaluates every proper subset
$S\subsetneq\mathcal D$ against the fixed background
$C_0\cup S$, including the empty set.
SD-ASR requires complete retrieval, a full-packet target
rate of at least $\tau=.50$, and proper-subset target rates
of at most $\epsilon=.10$.
Unless otherwise stated, confidence intervals use
10{,}000 benchmark-stratified query--target bootstrap resamples.
Defense thresholds
are frozen after tuning on disjoint development data.
Three blind annotators audit 240 returned packets
(pre-adjudication $\alpha=.74$--.86).
Two audit samples serve different purposes: 240 returned LENS
packets are assessed for semantic validity, while a separate
manifest of 240 fixed attempt IDs is shared across constructors
for all-attempt JTER evaluation.

\subsection{Main Results and Native Attacks}
\label{sec:main-results}

\begin{table*}[t]
\centering
\small
\setlength{\tabcolsep}{4.5pt}
\renewcommand{\arraystretch}{1.1}
\textbf{(a) LENS across benchmarks}\par\smallskip
\begin{tabular*}{\textwidth}{@{\extracolsep{\fill}}lrrcccr@{}}
\toprule
\textbf{Benchmark} & \textbf{Attempts} & \textbf{Returned} &
\textbf{Full@5} & \textbf{Any-hit@5} & \textbf{JRR@5} &
\textbf{TER \% [95\% CI]} \\
\midrule
HotpotQA          & 1{,}000 & 718   & 665   & .975 & .926 & 50.4 [48.2, 52.5] \\
2WikiMultihopQA   & 1{,}000 & 689   & 625   & .972 & .907 & 46.6 [44.4, 48.7] \\
MuSiQue           & 1{,}000 & 756   & 719   & .984 & .951 & 57.1 [54.9, 59.3] \\
Natural Questions & 1{,}000 & 594   & 529   & .960 & .891 & 34.3 [32.3, 36.3] \\
\midrule
\textbf{Macro}    & \textbf{4{,}000} & \textbf{2{,}757} & \textbf{2{,}538} & \textbf{.973} & \textbf{.919} & \textbf{47.1 [46.0, 48.2]} \\
\bottomrule
\end{tabular*}
\par\medskip
\textbf{(b) Native attack comparison}\par\smallskip
\begin{tabular*}{\textwidth}{@{\extracolsep{\fill}}lccccr@{}}
\toprule
\textbf{Attack (macro)} & \textbf{Docs} & \textbf{Ins.\ tok.} &
\textbf{JRR@5} & \textbf{TER \%} & \textbf{$\Delta$TER pts [CI]} \\
\midrule
PoisonedRAG        & 1   & 424  & .967 & 67.3 & $-20.2$ [$-22.1, -18.3$] \\
Semantic Chameleon & 2   & 765  & .892 & 47.4 & $-0.3$ [$-1.8, +1.2$] \\
Micro-Collaborative& 5   & 1{,}635 & .548 & 60.0 & $-12.9$ [$-14.8, -11.0$] \\
LENS               & 2/3 & 906  & .919 & 47.1 & reference \\
\bottomrule
\end{tabular*}
\caption{(a) Main results over all fixed attempts. TER averages target
output frequencies, with construction failures counted as zero; full
retrieval is reported separately. (b) Benchmark-macro comparison at each
attack's native document budget. $\Delta$TER is LENS minus the baseline
in percentage points, with paired 95\% confidence intervals.}
\label{tab:main}
\end{table*}

\begin{table}[tb]
\centering
\footnotesize
\setlength{\tabcolsep}{3pt}
\renewcommand{\arraystretch}{1.1}
\begin{tabularx}{\columnwidth}{@{}Lccc@{}}
\toprule
\textbf{Condition} & \multicolumn{2}{c}{\textbf{Out-of-scope}} & \textbf{In-scope} \\
\cmidrule(lr){2-3}
 & \textbf{TER} & \textbf{Correct} & \textbf{Correct} \\
\midrule
Sufficient clean evidence            & .012 & .927 & .918 \\
Length-matched neutral material       & .019 & .906 & .899 \\
Same LENS packet                      & \textbf{.708} & .231 & \textbf{.847} \\
Applicability check                  & .164 & .755 & .816 \\
\bottomrule
\end{tabularx}
\caption{Paired tasks on 240 instances using the same packets.
The packet raises out-of-scope TER by $+.696$ [.672, .721] over clean
evidence. In-scope correctness exceeds out-of-scope correctness by
$+.616$ [.588, .644]; brackets report paired 95\% CIs.}
\label{tab:paired}
\end{table}

\begin{table}[tb]
\centering
\footnotesize
\setlength{\tabcolsep}{3pt}
\renewcommand{\arraystretch}{1.1}
\begin{tabularx}{\columnwidth}{@{}Lccc@{}}
\toprule
\textbf{Constructor} & \textbf{Yield} & \textbf{\shortstack{TER\\\%}} &
\textbf{\shortstack{SemValid\\\%}} \\
\midrule
Generic repair            & .619 & 34.9 & 53.0 \\
Criteria-inspired adapt.  & .661 & 40.2 & 61.7 \\
LENS (full plan)          & .714 & 47.6 & 76.3 \\
\bottomrule
\end{tabularx}
\caption{Matched-objective construction with the same feedback and budget
($3{\times}1{,}000$ attempts). SemValid uses 300 audited returned
packets per constructor.}
\label{tab:matched}
\end{table}

\noindent\textbf{End-to-end target induction.}
LENS reaches 47.1\% TER over all 4{,}000 attempts (95\% CI [46.0, 48.2]),
with rates of 46.5--48.4\% across target generators
(Table~\ref{tab:main}). Among returned packets,
target-error rates are 39.3\% under partial retrieval and
71.9\% under complete retrieval.
Partial retrieval means that at least one, but not all, inserted
documents are retrieved; zero-hit cases remain in all-attempt TER.
Increasing $K$ from 3 to 8 raises JRR from .781 to .968 and all-attempt
TER from 42.0\% to 48.8\%.
Giving each rule equal weight within its benchmark yields
46.4\% TER, close to the original 47.1\%.

\noindent\textbf{Rule misuse and plain poisoning.}
The native-budget comparison measures each attack's target induction.
Semantic audits assess whether returned packets satisfy the
factual and scope conditions of Definition~\ref{def:misapplication}.
In a returned-packet audit of 240 LENS packets, excluding
construction failures, 75.8\% satisfy these semantic
requirements.
The externally checkable-rule stratum reaches 48.7\% TER.
Among the 168 externally checkable-rule packets in this audit,
132 pass semantic validation.
An independent generated-assertion control has zero semantic validity.
These results support the role of grounded rules in the
observed attack behavior.

\subsection{Matched Construction and Paired Tasks}
\label{sec:matched}

\noindent\textbf{Same content, different applicability.}
On paired tasks, the same packet induces 70.8\% target error outside the
rule's scope while preserving 84.7\% correct answers within it
(Table~\ref{tab:paired}). Length-matched neutral material leaves the
out-of-scope error at 1.9\%. Thus the packet remains useful for the task
its rule governs while misleading another task.
The paired questions test whether the same material is used
appropriately across tasks.
The clean and neutral conditions establish that the
out-of-scope error emerges when the scoped rule and its supporting facts
are introduced together.

\noindent\textbf{Matched construction.}
Across three 1{,}000-attempt runs with the same objective, diagnostics,
budget, and acceptance gate, LENS reaches 47.6\% TER versus
34.9\% for generic repair (Table~\ref{tab:matched}).
This 12.7-point improvement shows that structured planning
and repair strengthen target induction under matched conditions.
Because generic repair receives the same behavioral, retrieval,
and validity diagnostics, this comparison tests how the
constructor uses feedback. The explicit plan links each
correction to a rule, a scope condition, and a document role.

\begin{table}[tb]
\centering
\footnotesize
\setlength{\tabcolsep}{3pt}
\renewcommand{\arraystretch}{1.1}
\begin{tabularx}{\columnwidth}{@{}Lccc@{}}
\toprule
\textbf{Constructor} & \textbf{\shortstack{Returned\\/240}} &
\textbf{\shortstack{Audit-pool\\TER \%}} &
\textbf{\shortstack{JTER \% [95\% CI]}} \\
\midrule
Generic repair            & 141 & 36.5 & 20.7 [18.5, 22.9] \\
Criteria                 & 153 & 41.7 & 27.1 [24.7, 29.6] \\
LENS                     & 166 & 47.2 & \textbf{36.1} [33.7, 38.5] \\
\bottomrule
\end{tabularx}
\caption{Common audit list: 240 fixed attempts and three generators per
constructor. Returned packet--generator pairs receive 20 generations;
construction failures contribute zero to both rates. JTER joins each
instance--generator audit label to its target-output frequency.}
\label{tab:audit-pool}
\end{table}

\noindent\textbf{Joint semantic and behavioral evidence.}
A separate audit manifest fixes 240 query--target attempt IDs
before construction and shares them across constructors
(Table~\ref{tab:audit-pool}).
Unlike the returned-packet audit, this evaluation retains
construction failures. Independent validity judgments are
combined with frozen target responses to compute JTER.
LENS attains 36.1\% JTER [33.7, 38.5], versus 20.7\%
for generic repair and 27.1\% for the criteria-inspired
adaptation. The 15.4-point advantage over generic repair
shows that LENS remains stronger when both target induction
and independently audited validity are required.

\noindent\textbf{Component contributions.}
Fixed-budget ablations with independent validity audits identify
complementary contributions. Replacing typed
repair with generic repair costs 17.7 JTER points; removing role
assignment costs 12.6 and omitting the boundary field $\kappa$ costs 8.8.
Retrieval feedback improves exposure, while the semantic plan and typed
repair sustain valid misapplication.

\subsection{Mechanism: Intervention, Probe, and Re-segmentation}
\label{sec:mechanism}

\begin{figure}[tb]
\centering
\includegraphics[width=\columnwidth]{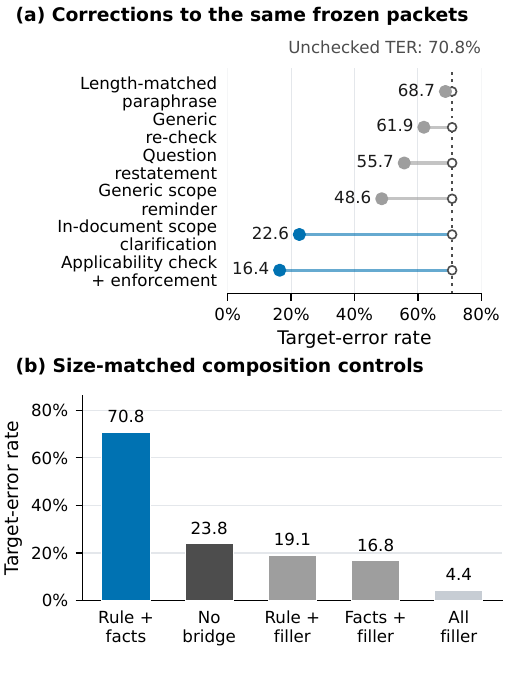}
\caption{(a) Each segment connects unchecked TER (open circle) to the
rate after one intervention (filled circle) on the same frozen packets.
Blue points target the scope relation. (b) Target-error rates for
size-matched composition controls.}
\label{fig:mechanism}
\end{figure}

\noindent\textbf{Applicability intervention.}
Sharpening the rule's scope while preserving facts, document count, and
order reduces TER from 70.8\% to 22.6\%. Checking applicability with the
documents frozen reduces it to 16.4\%, retaining 81.6\% in-scope accuracy
on the paired-task experiment. Equal-budget re-checking, clarification,
and generic scope reminders leave 48.6--61.9\% TER
(Figure~\ref{fig:mechanism}a). Scope is already visible in 95.4\% of
contexts; making it more salient reduces TER to 25.9\%. The effect
therefore depends on how the available scope is applied. Size-matched
rule/fact replacements reduce TER to 16.8--23.8\%
(Figure~\ref{fig:mechanism}b), linking the error to the planned bridge.

\noindent\textbf{Judgment and answering.}
Explicit applicability probes achieve .823--.850 balanced accuracy.
For 76.2\% of default responses matching the attacker target,
the same model correctly identifies the rule as inapplicable
when explicitly probed. This
within-instance contrast separates successful scope judgment from its
use during answering.
On the 181 packet--task pairs passing independent validity checks
in the original context, scope clarification reduces TER from
72.3\% to 24.8\%. Among 376 default responses matching the
attacker target, the same model correctly judges the rule
inapplicable in 281 cases (74.7\%) under explicit probing.

\noindent\textbf{Document boundaries.}
Merging or splitting fixed visible text changes TER by at most
2.7 points; re-indexing shorter or longer chunks changes it by at most
2.5. The mechanism persists across these
document boundaries.

\section{Checking Rule Applicability}
\label{sec:defense}

\begin{table}[t]
\centering
\footnotesize
\setlength{\tabcolsep}{2.5pt}
\renewcommand{\arraystretch}{1.12}
\begin{tabularx}{\columnwidth}{@{}Lccccc@{}}
\toprule
\textbf{Check} & \textbf{\shortstack{TER\\\%}} & \textbf{Gold} &
\textbf{\shortstack{Legit.\\acc.}} & \textbf{\shortstack{Clean\\acc.}} &
\textbf{\shortstack{Total cost\\calls / tok.}} \\
\midrule
None                     & 70.8 & .231 & .851 & .724 & 1.0 / 2.9k \\
Passage screening        & 67.6 & .248 & .829 & .711 & 2.0 / 4.8k \\
Generic re-check         & 61.9 & .307 & .833 & .715 & 2.0 / 4.7k \\
Fact-checking            & 65.1 & .271 & .816 & .704 & 2.0 / 5.1k \\
Set-level judge          & 40.1 & .505 & .761 & .682 & 2.0 / 5.3k \\
Task-alignment           & 31.2 & .584 & .795 & .695 & 2.0 / 4.9k \\
\textbf{Applicability audit} & \textbf{16.4} & \textbf{.733} & \textbf{.821} & \textbf{.708} & 2.0 / 5.0k \\
\bottomrule
\end{tabularx}
\caption{Matched-information, matched-resource-cap comparison
(720 evaluations). Total costs include extraction, judgment, and final
generation. Legitimate-task results use the paired-task pool with
independent generation sampling.}
\label{tab:defense}
\end{table}

\begin{table*}[t]
\centering
\small
\setlength{\tabcolsep}{3pt}
\renewcommand{\arraystretch}{1.1}
\begin{tabular*}{\textwidth}{@{\extracolsep{\fill}}lccccccc@{}}
\toprule
\textbf{Attack} & \textbf{None} & \textbf{RobustRAG} &
\textbf{\shortstack{Reliability\\RAG}} & \textbf{SeCon-RAG} & \textbf{TrustRAG} &
\textbf{PRA-RAG} & \textbf{\shortstack{Audit\\(ours)}} \\
\midrule
PoisonedRAG              & 67.3 & 22.5 & 31.4 & 23.6 & 28.2 & \textbf{12.1} & 28.9 \\
Semantic Chameleon       & 47.4 & 24.9 & 27.8 & 20.7 & 25.8 & 14.6 & 21.8 \\
Micro-Collaborative      & 60.0 & 23.5 & 32.4 & 24.8 & 29.9 & 16.9 & 26.7 \\
Criteria-inspired adapt. & 41.8 & 27.1 & 29.6 & 24.6 & 27.3 & 16.4 & \textbf{13.7} \\
LENS                     & 47.1 & \textbf{32.6} & \textbf{37.4} & \textbf{34.4} & \textbf{35.9} & \textbf{21.8} & \textbf{11.9} \\
\bottomrule
\end{tabular*}
\caption{Target-error rate (\%, all attempts) of each attack under each
defense. LENS retains the highest TER under each published defense;
applicability auditing most reduces the scope-misuse attacks.}
\label{tab:defense-matrix}
\end{table*}

\begin{figure}[!htb]
\centering
\includegraphics[width=\columnwidth]{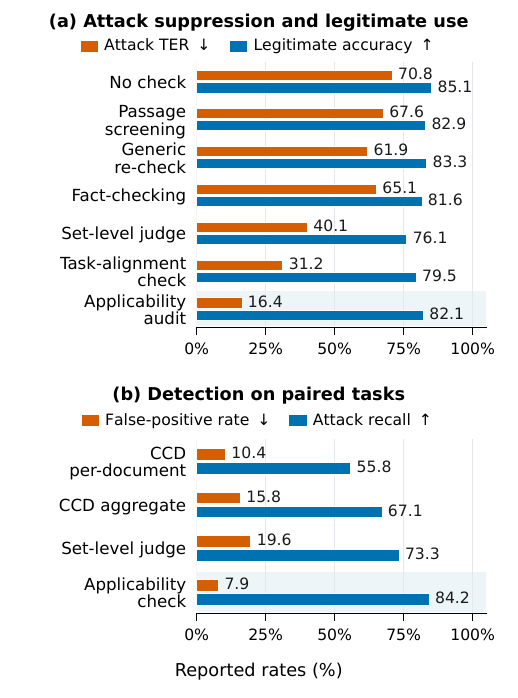}
\caption{Checks with matched information and resource caps.
(a) Attack TER and legitimate-task accuracy. (b) Attack recall and
legitimate-task false positives at frozen operating points.}
\label{fig:defense}
\end{figure}

\noindent\textbf{Applicability audit.}
The first call extracts task requirements from the query and
application settings, identifies candidate rules and their
stated scopes in the retrieved context, and judges applicability
with supporting evidence positions. The second call receives
these verdicts and is instructed to use only rules marked
applicable. We implement the applicability gate through prompting and
measure enforcement errors separately.
Source authorization is determined by the user or application,
not by retrieved documents.
Documents remain available for legitimate uses; neither
call receives reference answers, targets, or document labels. For AOSP,
the audit excludes the kernel-patch provision from the majority-license
answer while retaining it for the paired patch question. Total cost is
2.0 calls and 5.0k tokens, an increment of 1.0 call and 2.1k tokens over
unchecked answering.

\noindent\textbf{Safety and utility.}
Table~\ref{tab:defense} compares checks using the same query,
retrieved context, and application constraints, with matched
resource caps that include extraction costs.
Its legitimate-task pool is the same 240-case pool as
Table~\ref{tab:paired}, evaluated with an independent generation batch.
Applicability auditing reduces TER to 16.4\%, versus 31.2\% for
task-alignment checking and 40.1\% for a set-level judge. Legitimate-task
accuracy remains 82.1\% (85.1\% unchecked), and clean accuracy remains
70.8\% (72.4\% unchecked). On paired-task detection, it recalls 84.2\%
of attacks at 7.9\% legitimate false positives; the set-level judge
reaches 73.3\% at 19.6\% (Figure~\ref{fig:defense}).
Enforcement is essential: judging applicability without constraining
final generation leaves 52.7\% TER, compared with 16.4\% for the complete
audit. A correct check must govern the final answer.

\noindent\textbf{Published defenses.}
Across the full attempt manifest, LENS retains the highest TER under
each published defense (21.8--37.4\%; Table~\ref{tab:defense-matrix}).
Applicability auditing reduces LENS to 11.9\% and the criteria-inspired
attack to 13.7\%, directly addressing their shared misuse of answer
criteria. Its benefit complements defenses against explicit false
statements: RobustRAG reaches 22.5\% on PoisonedRAG, compared with
28.9\% for applicability auditing.

\noindent\textbf{Adaptive attacks.}
On a separate 1{,}000-attempt adaptive manifest, attackers target task
extraction, rule--source binding, the scope--task bridge, and verdict
enforcement, or combine these strategies through local selection.
They receive three additional actions and 0.30M tokens, for total caps of
18 actions and 1.35M tokens per attempt. The non-adaptive baseline
and all adaptive strategies share these caps.
Under this expanded budget,
the strongest combined strategy increases defended TER from the
same-manifest non-adaptive 12.1\% to 24.1\%; semantic validity changes
from 77/100 to 69/100 audited packets. Among 160 audited residual
failures, the most common first errors arise in rule/scope extraction or binding
(26.3\%), applicability judgment
(24.4\%), verdict enforcement (21.9\%), and task extraction (19.4\%).
On the same 720 packet--generator pairs, replacing automatic
task and rule/scope extraction and applicability judgments with
audited labels, while leaving enforcement and generation automatic,
reduces TER from 16.4\% to 5.4\%.
These results identify where the practical audit falls short of
the conditions in Proposition~\ref{prop:scoped}.

\section{Related Work}
\label{sec:related}

\paragraph{Corpus poisoning and cross-document coordination.}
PoisonedRAG constructs answer-steering documents \citep{zou2025poisonedrag},
while GASLITE and Joint-GCG optimize retrieval-side and joint objectives
\citep{gaslite2025,wang2026jointgcg}. Semantic Chameleon,
Micro-Collaborative Poisoning, and COMA demonstrate cross-document
influence \citep{semanticchameleon2026,pereira2026micro,
gondhalekar2026coma}. LENS instead studies the misapplication of a grounded rule whose
scope remains explicit, using paired questions to test whether
the same evidence supports legitimate in-scope answers.

\paragraph{Standards, criteria, and task definition.}
Prompt-level attacks manipulate judgment criteria
\citep{criteriaattack2026}, and contextual-integrity work studies how
untrusted context shifts norms and authorization scopes \citep{ci2026}.
Our focus is whether a valid rule can steer an answer beyond its
explicit scope despite the presence of correct evidence.
Paired tasks and scope interventions examine this misuse, while
separate probes distinguish explicit applicability judgments
from default answer behavior.

\paragraph{Robust retrieval-augmented generation.}
RobustRAG, ReliabilityRAG, SeCon-RAG, TrustRAG, and PRA-RAG use evidence
isolation, reliability checks, or aggregation
\citep{xiang2024robustrag,reliabilityrag2025,seconrag2025,trustrag2025,
tan-etal-2026-pra}; RAG2RAG adds a trained risk assessor
\citep{liang2026rag2rag}. These methods motivate our matched-information
comparison of content-focused checks with explicit applicability checking
and enforcement (\S\ref{sec:defense}), reporting legitimate-task utility
alongside attack reduction.

\section{Conclusion}
\label{sec:conclusion}

We identified rule-scope misapplication as a RAG security failure in which
true facts and a locally valid rule support an incorrect answer because
the rule is applied beyond its stated scope. LENS constructs auditable
instances, paired tasks and interventions isolate the mechanism, and an
applicability audit reduces target errors while preserving legitimate
evidence use. Explicit probes show that models can correctly judge a
rule inapplicable yet still produce the target error when answering
normally. These findings motivate checking rule applicability and
enforcing the judgment during answer generation.

\section*{Limitations}
\label{sec:limitations}

Experiments cover single-round, text-based RAG with fixed queries and
documented rule scopes. Reported attack rates are conditional on corpus
admission and selection of clean-answerable, semantically screened
questions. Generalization to ambiguous tasks and interactive or
multimodal retrieval requires further evaluation. Applicability auditing
relies on trusted task requirements and correct rule extraction,
judgment, and enforcement; adaptive attacks can exploit errors in these
steps. Human audits are subject to sampling and annotation uncertainty.

\paragraph{Ethical considerations.}
Evaluation uses benchmark corpora and controlled ingestion replays to
study and mitigate rule misuse. The attack procedure could also be used
to manipulate deployed assistants and mislead users. Its use should be
limited to systems for which testing is authorized.

\begingroup
\setlength{\emergencystretch}{2em}
\bibliography{slcf_refs}
\endgroup

\end{document}